\documentclass[graybox]{svmult}
\usepackage{amsmath,amsfonts,mathrsfs,amssymb}
\usepackage{mathptmx}       
\usepackage{helvet}         
\usepackage{courier}        
\usepackage{type1cm}        
\usepackage{makeidx}         
\usepackage{graphicx}        
\usepackage{multicol}        
\usepackage[bottom]{footmisc}
\usepackage{epsfig}

\makeindex             

\newcommand\mr{\mathscr}

\begin{document}

\title*{Dynamical holographic QCD model: resembling renormalization group
from ultraviolet to infrared}
\titlerunning{Dynamical holographic QCD model: resembling RG from UV to IR}
\author{Danning Li and Mei Huang}
\institute{Danning Li \at Institute of High Energy Physics, Chinese Academy of Sciences,
Beijing 100049 \\ \email{lidn@mail.ihep.ac.cn}
\and Mei Huang (speaker) \at Institute of High Energy Physics, Chinese Academy of Sciences, Beijing 100049 \\
\email{huangm@mail.ihep.ac.cn}}
%
%
\maketitle

\abstract{Resembling the renormalization group from ultraviolet (UV) to infrared (IR),
we construct a dynamical holographic model in the graviton-dilaton-scalar
framework, where the dilaton background field $\Phi$ and scalar field $X$ are
responsible for the gluodynamics and chiral dynamics, respectively. At the UV boundary,
the dilaton field is dual to the dimension-4 gluon operator, and the scalar field is
dual to the dimension-3 quark-antiquark operator. The metric structure at IR is
automatically deformed by the nonperturbative gluon condensation and chiral
condensation in the vacuum. The produced scalar glueball spectra in the
graviton-dilaton framework agree well with lattice data, and the light-flavor
meson spectra generated in the graviton-dilaton-scalar framework are in well
agreement with experimental data. Both the chiral symmetry breaking and linear
confinement are realized in this dynamical holographic QCD model.
The necessary condition for the existence of linear quark potential is
discussed, and the pion form factor is also investigated in the dynamical
hQCD model.}

\section{Introduction}
\label{sec:1}

Quantum chromodynamics (QCD) is accepted as the fundamental theory of
the strong interaction. In the ultraviolet (UV) or weak coupling regime of
QCD, the perturbative calculations agree well with experiment. However, in the
infrared (IR) regime, the description of QCD vacuum as well as hadron properties
and processes in terms of quark and gluon still remains as outstanding challenge
in the formulation of QCD as a local quantum field theory.

\begin{figure}[!htb]
\includegraphics[scale=.5]{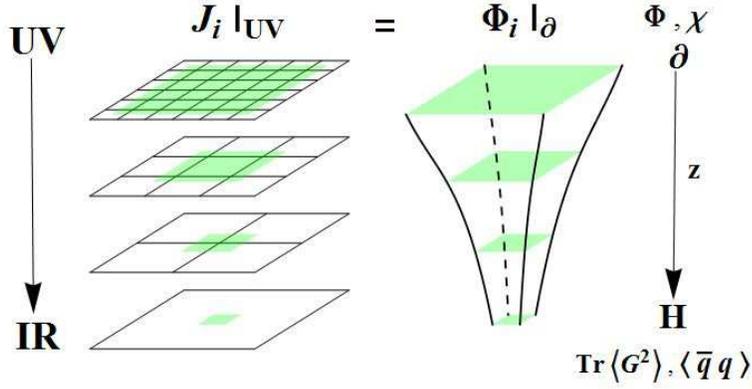}
\caption{Duality between $d$-dimension QFT and $d+1$-dimension gravity as shown
in \cite{Adams:2012th} (Left-hand side). Dynamical holographic QCD model resembles
RG from UV to IR (Right-hand side): at UV boundary the dilaton bulk field $\Phi(z)$
and scalar field $X(z)$ are dual to the dimension-4 gluon operator and dimension-3
quark-antiquark operator, which develop condensates at IR. }
\label{fig:RGflow}
\end{figure}

In order to derive the
low-energy hadron physics and understand the deep-infrared sector of QCD from first
principle, various non-perturbative methods have been employed, in particular
lattice QCD, Dyson-Schwinger equations (DSEs), and functional renormalization
group equations (FRGs). In recent decades, an entirely new method based on the
anti-de Sitter/conformal field theory (AdS/CFT) correspondence and the conjecture
of the gravity/gauge duality \cite{Maldacena:1997re,Gubser:1998bc,Witten:1998qj}
provides a revolutionary method to tackle the problem of strongly coupled gauge theories.
Though the original discovery of holographic duality requires supersymmetry and
conformality, the holographic duality has been widely used in investigating hadron
physics \cite{Karch:2006pv,Csaki:2006ji,Gherghetta-Kapusta-Kelley,YLWu},
strongly coupled quark gluon plasma and condensed matter.
It is widely believed that the duality between the quantum field theory and quantum gravity
is an unproven but true fact. In general, holography relates quantum field theory (QFT) in
d-dimensions to quantum gravity in (d + 1)-dimensions, with the gravitational description
becoming classical when the QFT is strongly-coupled. The extra dimension
can be interpreted as an energy scale or renormalization group (RG) flow in the QFT
\cite{Adams:2012th} as shown in Fig.\ref{fig:RGflow}.

In this talk, we introduce our recently developed dynamical holographic QCD model \cite{DhQCD},
which resembles the renormalization group from ultraviolet (UV) to infrared (IR).
The dynamical holographic model is constructed in the graviton-dilaton-scalar
framework, where the dilaton background field $\Phi(z)$ and scalar field $X(z)$ are
responsible for the gluodynamics and chiral dynamics, respectively. At the UV boundary,
the dilaton field $\Phi(z)$ is dual to the dimension-4 gluon operator, and the scalar
field $X(z)$ is dual to the dimension-3 quark-antiquark operator. The metric structure
at IR is automatically deformed by the nonperturbative gluon condensation and chiral
condensation in the vacuum.  In Fig.\ref{fig:RGflow}, we show the dynamical holographic
QCD model, which resembles the renormalization group from UV to IR.

\section{Pure gluon system: Graviton-dilaton framework}
\label{sec:2}

For the pure gluon system, we construct the quenched dynamical holographic QCD model
in the graviton-dilaton framework by introducing one scalar dilaton
field $\Phi(z)$ in the bulk. The 5D graviton-dilaton coupled action in the string frame
is given below:
\begin{eqnarray}\label{action-graviton-dilaton}
 S_G=\frac{1}{16\pi G_5}\int
 d^5x\sqrt{g_s}e^{-2\Phi}\left(R_s+4\partial_M\Phi\partial^M\Phi-V^s_G(\Phi)\right).
\end{eqnarray}
Where $G_5$ is the 5D Newton constant, $g_s$, $\Phi$ and $V_G^s$ are the 5D
metric, the dilaton field and dilaton potential in the string frame, respectively.
The metric ansatz is often chosen to be
\begin{eqnarray}\label{metric-ansatz}
ds^2=b_s^2(z)(dz^2+\eta_{\mu\nu}dx^\mu dx^\nu), ~ ~ b_s(z)\equiv e^{A_s(z)}.
\end{eqnarray}

To avoid the gauge non-invariant problem and to meet the requirement of gauge/gravity duality,
we take the dilaton field in the form of
\begin{equation}
\Phi(z)=\mu_G^2z^2\tanh(\mu_{G^2}^4z^2/\mu_G^2).
\label{mixed-dilaton}
\end{equation}
In this way, the dilaton field at UV behaves
$\Phi(z)\overset{z\rightarrow0}{\rightarrow} \mu_{G^2}^4 z^4$,
and is dual to the dimension-4 gauge invariant gluon operator ${\rm Tr} G^2 $,
while at IR it takes the quadratic form
$\Phi(z)\overset{z\rightarrow\infty}{\rightarrow} \mu_G^2 z^2$. By  self-consistently
solving the Einstein equations, the metric
structure will be automatically deformed at IR by the dilaton background field, for
details, please refer to \cite{DhQCD}.

\begin{figure}[!htb]
\sidecaption
\includegraphics[scale=0.75]{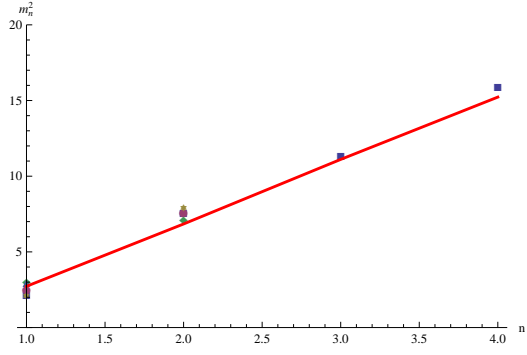}
\caption{The scalar glueball spectra for the dilaton field
$\Phi(z)=\mu_G^2z^2\tanh(\mu_{G^2}^4z^2/\mu_G^2)$
with $\mu_G=\mu_{G^2}=1 {\rm GeV}$. The dots are lattice data taken from
\cite{glueball-lattice}.}
\label{z4-z2glueball}
\end{figure}

We assume the glueball can be excited from the QCD vacuum described by the
quenched dynamical holographic model, and
the 5D action for the scalar glueball $\mathscr{G}(x,z)$ in the string frame
takes the form as
\begin{eqnarray}
S_{\mathscr{G}}=\int d^5 x \sqrt{g_s}\frac{1}{2}e^{-\Phi}\big[ \partial_M \mathscr{G}\partial^M
\mathscr{G}+M_{\mathscr{G},5}^2 \mathscr{G}^2\big].
\end{eqnarray}

The Equation of motion for $\mathscr{G}$ has the form of
\begin{eqnarray}
-e^{-(3A_s-\Phi)}\partial_z(e^{3A_s-\Phi}\partial_z\mr{G}_n)=m_{\mathscr{G},n}^2 \mathscr{G}_n.
\end{eqnarray}
After the transformation $\mr{G}_n \rightarrow e^{-\frac{1}{2}(3A_s-\Phi)}\mathscr{G}_n$,
we get the schrodinger like equation of motion for the scalar glueball
\begin{eqnarray}
-\mathscr{G}_n^{''}+V_{\mathscr{G}} \mathscr{G}_n=
m_{\mathscr{G},n}^2 \mathscr{G}_n,
\label{EOM-glueball}
\end{eqnarray}
with the 5D effective schrodinger potential
\begin{equation}
V_{\mathscr{G}}=\frac{3A_s^{''}-\Phi^{''}}{2}+\frac{(3A_s^{'}-\Phi^{'})^2}{4}.
\label{potential-glueball}
\end{equation}

Then from Eq. (\ref{EOM-glueball}), we can solve the scalar glueball spectra
and the result is shown in Fig.\ref{z4-z2glueball}. It is a surprising result
that if one self-consistently solves the metric background under the dynamical
dilaton field, it gives the correct ground state and at the same time gives the
correct Regge slope.

\begin{figure}[!htb]
\epsfxsize=6.5 cm \epsfysize=6.5 cm \epsfbox{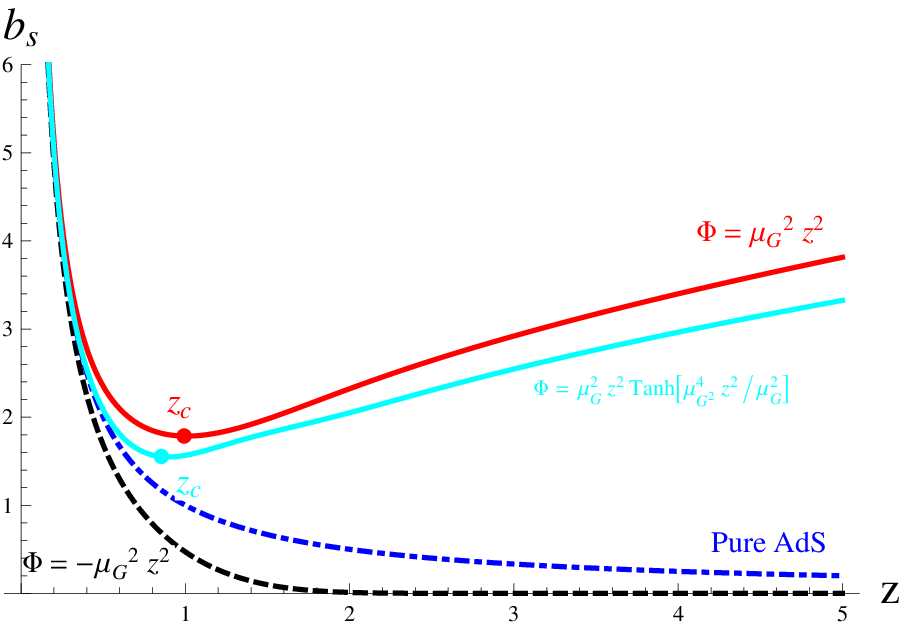} \hspace*{0.1cm}
\epsfxsize=6.5 cm \epsfysize=6.5 cm \epsfbox{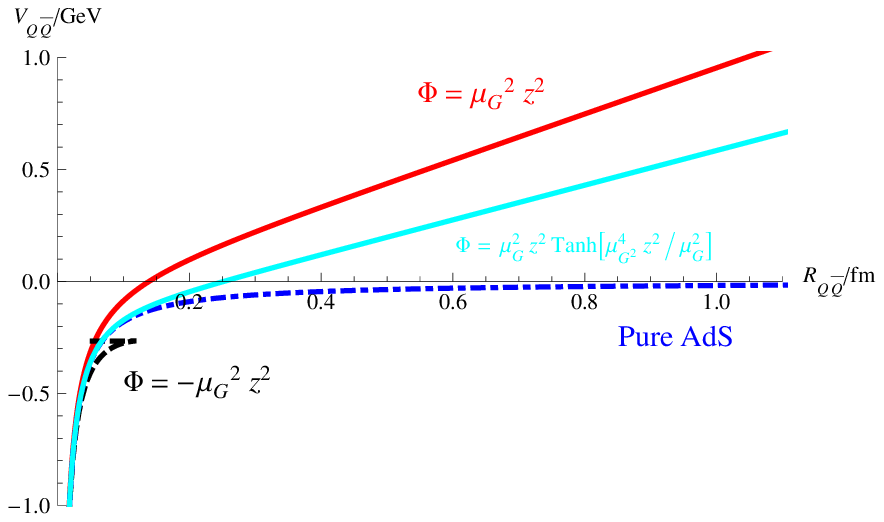} \vskip -0.05cm
\hskip 0.15 cm
\textbf{( $b_s$ ) } \hskip 6.5 cm \textbf{( $V_{Q\bar Q} $ )} \\
\caption{The metric structure $b_s(z)=e^{A_s(z)}$ as functions of $z$ (feft-hand side),
and the quenched quark potential result $V_{Q\bar Q} $ as functions of
$R_{Q\bar Q}$ (right-hand side)
corresponding to $\Phi=\mu_G^2z^2$ (red solid line), $\Phi=-\mu_G^2z^2$ (black dashed line),
and $\Phi=\mu_G^2z^2\tanh(\mu_{G^2}^4z^2/\mu_G^2)$ (green solid line), respectively.
The blue dash-dotted line stands for the pure ${\rm AdS}_5$ case. $\mu_G=1 {\rm GeV}$ has been taken for numerical calculation.}
\label{quenched-bs}
\end{figure}

Following the standard procedure, the heavy quark potential $V_{Q\bar Q}$ and the
interquark distance $R_{Q\bar Q}$ can be worked out.
We also find the necessary condition for the linear quark potential:
There exists a point $z_c$, at which
$b_s^{'}(z_c)\rightarrow 0, b_s(z_c)\rightarrow const$,
then one can obtain the string tension
\begin{eqnarray}
\sigma_s \propto \frac{V_{Q\bar Q}(z_0)}{R_{\bar{q}q}(z_0)}\overset{z_0\rightarrow z_c}{\longrightarrow} \frac{L^2}{2\pi \alpha_p} b_s^2(z_c). \label{stringtension-g}
\end{eqnarray}
Where $\alpha_p$ is the 5D string tension. From the left-hand figure in
Fig.\ref{quenched-bs}, we can see that only for
the case of positive dilaton background $\Phi=\mu_G^2z^2$ and $\Phi=\mu_G^2z^2\tanh(\mu_{G^2}^4z^2/\mu_G^2)$, the metric has a
minimum point $z_c$. Correspondingly, the quark-antiquark potential
indeed shows a linear part for positive quadratic dilaton background $\Phi=\mu_G^2z^2$ and
for $\Phi=\mu_G^2z^2\tanh(\mu_{G^2}^4z^2/\mu_G^2)$ as shown in right-hand figure in
Fig.\ref{quenched-bs}. While for the pure ${\rm AdS}_5$ case
as well as for the dynamical soft-wall model with negative dilaton background field
$\Phi=-\mu_G^2z^2$, there doesn't exist a $z_c$ where $b_s^{'}(z_c)\rightarrow 0$, and
correspondingly the heavy quark potential does not show a linear behavior at large $z$.

\section{Dynamical holographic QCD model for meson spectra}
\label{sec:3}

We then add light flavors in terms of meson fields on the gluodynamical background.
The total 5D action for the graviton-dilaton-scalar system takes the following form:
\begin{eqnarray}
 S=S_G + \frac{N_f}{N_c} S_{KKSS},
\end{eqnarray}
with
\begin{eqnarray}
 S_G=&&\frac{1}{16\pi G_5}\int
 d^5x\sqrt{g_s}e^{-2\Phi}\big(R+4\partial_M\Phi\partial^M\Phi-V_G(\Phi)\big), \\
 S_{KKSS}=&&-\int d^5x
 \sqrt{g_s}e^{-\Phi}Tr(|DX|^2+V_X(X^+X, \Phi)+\frac{1}{4g_5^2}(F_L^2+F_R^2)).
\end{eqnarray}

\begin{figure}[!htb]
\epsfxsize=6.5 cm \epsfysize=6.5 cm \epsfbox{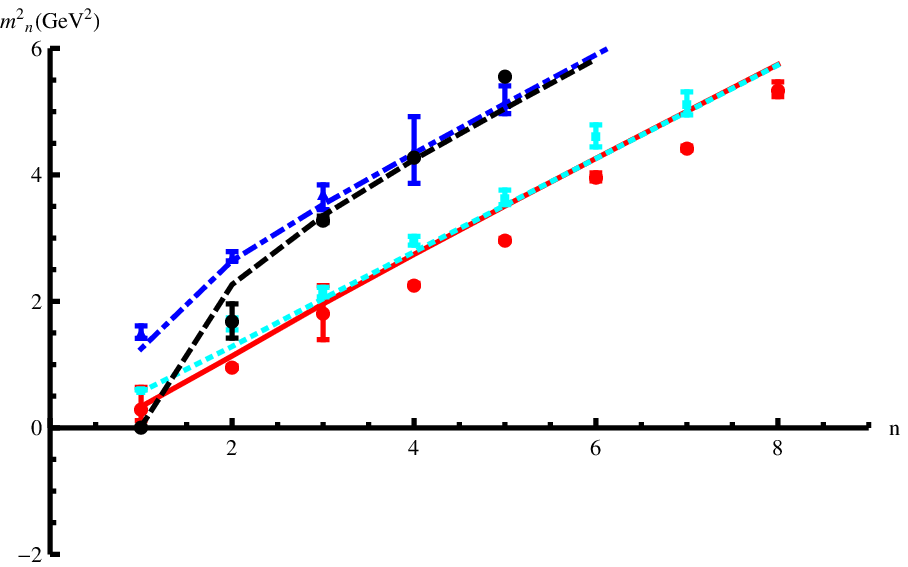} \hspace*{0.1cm}
\epsfxsize=6.5 cm \epsfysize=6.5 cm \epsfbox{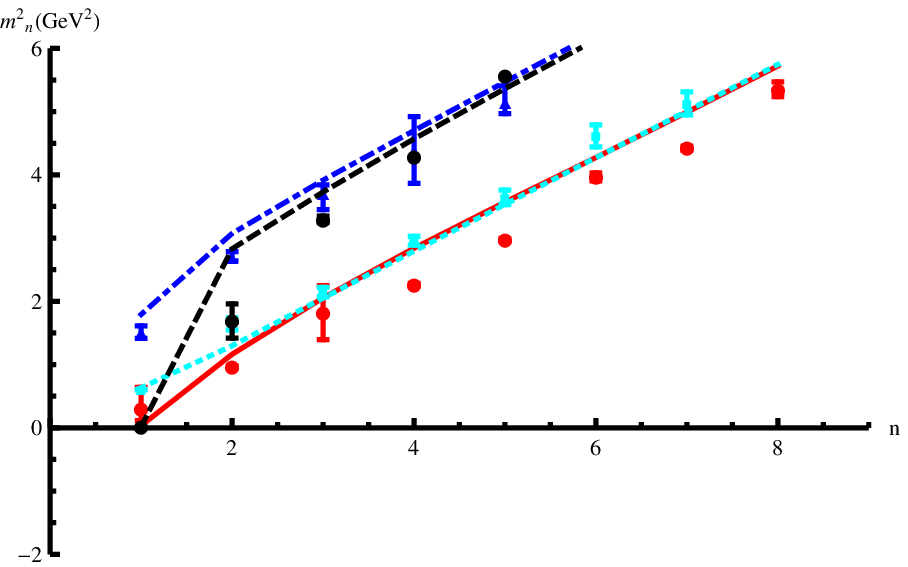} \vskip -0.05cm
\hskip 0.15 cm
\textbf{( Mod A ) } \hskip 6.5 cm \textbf{( Mod B )} \\
\caption{Meson spectra in the dynamical soft-wall model with two sets
of parameters in Table \ref{parameters} comparing with experimental data. The
red and black lines are for scalars and pseudoscalars, the green and blue
lines are for vectors and axial-vectors. }
\label{allmassespic}
\end{figure}

\begin{figure}[!htb]
\epsfxsize=6.5 cm \epsfysize=6.5 cm \epsfbox{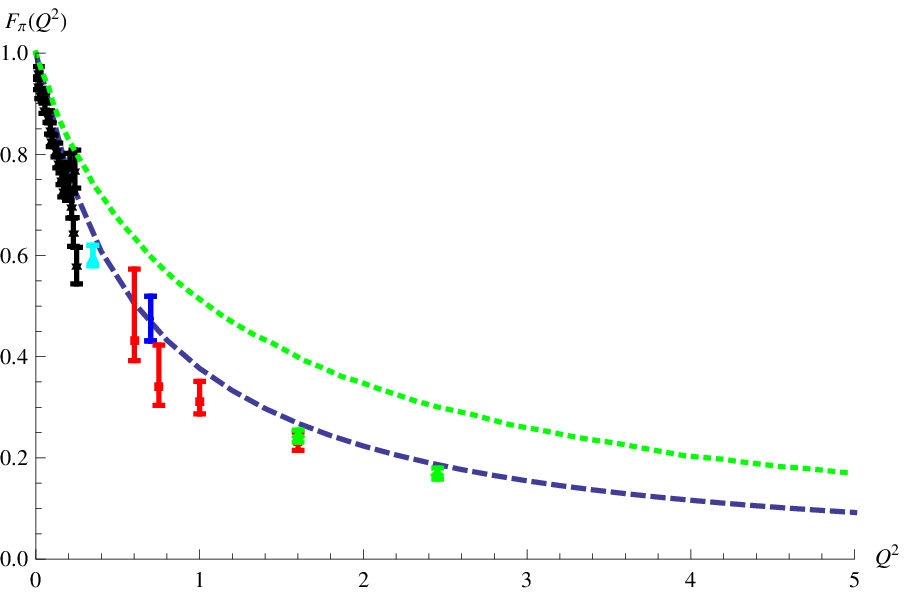} \hspace*{0.1cm}
\epsfxsize=6.5 cm \epsfysize=6.5 cm \epsfbox{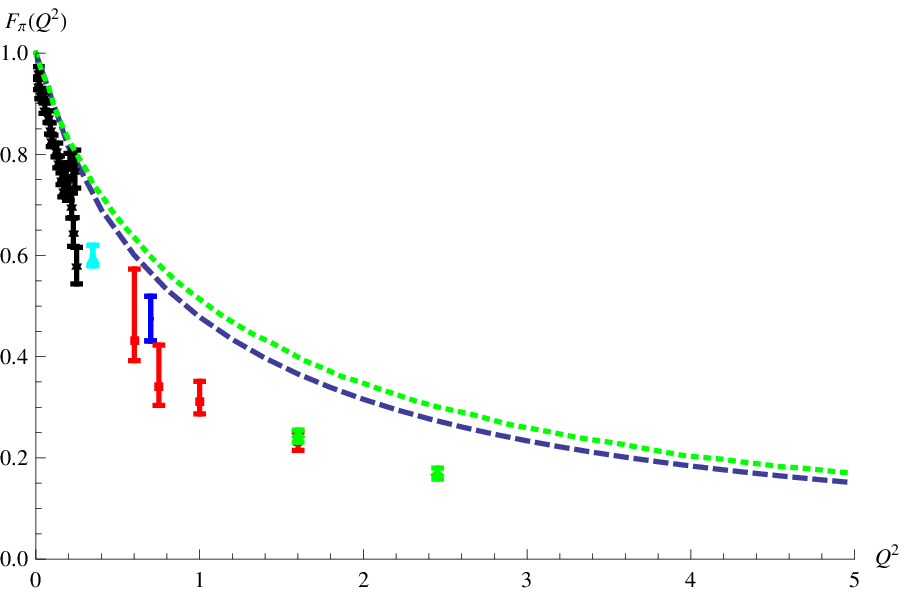} \vskip -0.05cm
\hskip 0.15 cm
\textbf{( Mod A ) } \hskip 6.5 cm \textbf{( Mod B )} \\
\caption[]{$F_\pi(Q^2)$ as function of $Q^2$ for Mod A and B defined in Table \ref{parameters}
and compared with experimental data. The blue dashed lines are the prediction in our model, and the green dotted line is the original soft-wall model results taken from Ref.\cite{Kwee:2007}.} \label{formfactors}
\end{figure}

In the vacuum, it is assumed that there are both gluon condensate and chiral condensate.
The dilaton background field $\Phi$ is supposed to be dual to some kind of gluodynamics
in QCD vacuum. We take the dilaton background field  $\Phi(z)=\mu_G^2z^2\tanh(\mu_{G^2}^4z^2/\mu_G^2)$. The scalar field $X(z)$ is
dual to dimension-3 quark-antiquark operator, and $\chi(z)$ is the vacuum expectation
value (VEV) of the scalar field $X(z)$. For detailed analysis please refer to \cite{DhQCD}.
The equations of motion of the vector, axial-vector, scalar and pseudo-scalar mesons
take the form of:
\begin{eqnarray}
-\rho_n^{''}+V_{\rho} \rho_n&=&m_n^2 \rho_n, \\
-a_n^{''}+ V_a a_n&=& m_n^2 a_n, \\
-s_n^{''}+V_s s_n&=&m_n^2 s_n, \\
 -\pi_n''+V_{\pi,\varphi} \pi_n & = & m_n^2(\pi_n-e^{A_s}\chi\varphi_n), \nonumber\\
 -\varphi_n''+V_{\varphi} \varphi_n & = & g_5^2 e^{A_s}\chi(\pi_n-e^{A_s}\chi\varphi_n).
\end{eqnarray}
with schrodinger like potentials
\begin{eqnarray}
V_{\rho}&=& \frac{A_s^{'}-\Phi^{'}}{2}+\frac{(A_s^{'}-\Phi^{'})^2}{4}, \\
V_a&=& \frac{A_s^{'}-\Phi^{'}}{2}+\frac{(A_s^{'}-\Phi^{'})^2}{4}+g_5^2 e^{2A_s} \chi^{2},\\
V_s&=& \frac{3A_s^{''}-\phi^{''}}{2}+\frac{(3A_s^{'}-\phi^{'})^2}{4}
            +e^{2A_s}V_{C,\chi\chi}(\chi,\Phi), \\
V_{\pi,\varphi}&=& \frac{3A_s^{''}-\Phi^{''}+2\chi^{''}/\chi-2\chi^{'2}/\chi^2}{2}
           +\frac{(3A_s^{'}-\Phi^{'}+2\chi^{'}/\chi)^2}{4}, \\
V_{\varphi}&=& \frac{A_s^{''}-\Phi^{''}}{2}+\frac{(A_s^{'}-\Phi^{'})^2}{4}.
\end{eqnarray}

For our numerical calculations, we take two sets of parameters in Table \ref{parameters}.
The parameters in Mod A has a smaller chiral condensate, which gives a smaller pion decay
constant $f_{\pi}=65.7 {\rm MeV}$, and the parameters in Mod B has a larger chiral condensate,
which gives a reasonable pion decay constant $f_{\pi}=87.4 {\rm MeV}$.

\begin{table}
\begin{center}
\begin{tabular}{cccccccc}
\hline\hline
  ~  &  $G_5/L^3$  &   $m_q$~(MeV)  &  $\sigma^{1/3}~(MeV)$ & $\mu_G=\mu_{G^2}$  \\   \hline
Mod~A   & 0.75        & 8.4     &165 & 0.43    \\
  Mod~B &  0.75       &6.2      & 226 & 0.43      \\
\hline\hline
\end{tabular}
\caption{Two sets of parameters.}
\label{parameters}
\end{center}
\end{table}

The meson spectra and pion form factor are shown in Fig.\ref{allmassespic} and Fig. \ref{formfactors}. It is observed that from Fig.\ref{allmassespic} that in our
graviton-dilaton-scalar system, with two sets of parameters, the generated meson
spectra agree well with experimental data. For the pion form factor, it is found
that with parameters set A used with a smaller chiral condensate,
the produced pion form factor matches the experimental data much better, however,
the produced pion decay constant is much smaller than experimental data.
With parameters in set B corresponding to a larger chiral condensate, one can produce better result for pion decay constant, but the results on pion form factor are worse.

\section{Discussion and summary}
\label{sec-summary}

In this work, we construct a quenched dynamical holographic QCD (hQCD) model
in the graviton-dilaton framework for the pure gluon system, and develop a dynamical
hQCD model for the two flavor system in the graviton-dilaton-scalar framework
by adding light flavors on the gluodynamical background.
The dynamical holographic model resembles the renormalization group from ultraviolet
(UV) to infrared (IR). The dilaton background field $\Phi$ and scalar field $X$ are
responsible for the gluodynamics and chiral dynamics, respectively. At the UV boundary,
the dilaton field is dual to the dimension-4 gluon operator, and the scalar field is
dual to the dimension-3 quark-antiquark operator. The metric structure at IR is
automatically deformed by the nonperturbative gluon condensation and chiral
condensation in the vacuum.
The produced scalar glueball spectra in the graviton-dilaton framework
agree well with lattice data, and the light-flavor
meson spectra generated in the graviton-dilaton-scalar framework are in well
agreement with experimental data. Both the chiral symmetry breaking and linear
confinement are realized in the dynamical holographic QCD model.

We also give a necessary condition
for the existence of linear quark potential from the metric structure, and we show
that in the graviton-dilaton framework, a negative quadratic dilaton background
field cannot produce the linear quark potential.

The pion form factor is also investigated in the dynamical
hQCD model. It is found that with smaller chiral condensate,
the produced pion form factor matches the experimental data much better, however,
the produced pion decay constant is much smaller than experimental data. With
larger chiral condensate, one can produce better result for pion decay constant,
but the result on pion form factor is worse.

\begin{acknowledgement}
This work is supported by the NSFC under Grant
Nos. 11175251 and 11275213, DFG and NSFC (CRC 110),
CAS key project KJCX2-EW-N01, K.C.Wong Education Foundation, and
Youth Innovation Promotion Association of CAS.
\end{acknowledgement}

\end{document}